\documentclass[epj]{webofc}
\usepackage[utf8]{inputenc}
\usepackage[varg]{txfonts}   
\usepackage{booktabs}
\usepackage{braket}
\usepackage{xcolor}
\definecolor{darkred}{rgb}{0.4,0.0,0.0}
\definecolor{darkgreen}{rgb}{0.0,0.4,0.0}
\definecolor{darkblue}{rgb}{0.0,0.0,0.4}
\usepackage[bookmarks,linktocpage,colorlinks,
    linkcolor = darkred,
    urlcolor  = darkblue,
    citecolor = darkgreen]{hyperref}
\usepackage{subfigure}
\wocname{EPJ Web of Conferences}
\woctitle{Lattice2017}
\newcommand{\definedby}{:=}
\newcommand{\projector}[1]{\Gamma_{\!#1}}
\newcommand{\proj}{\projector{\vect{p}}}
\newcommand{\identity}[0]{\mathbb{I}}
\newcommand{\vect}[1]{\mathbf{#1}}

\begin{document}
%
\selectlanguage{english}
\title{%
Structure of the Nucleon and its Excitations
}
\author{%
\firstname{Waseem} \lastname{Kamleh}\inst{1} \and
\firstname{Derek} \lastname{Leinweber}\inst{1}\fnsep\thanks{Speaker, \email{derek.leinweber@adelaide.edu.au}} \and
\firstname{Zhan-wei} \lastname{Liu}\inst{1,2} \and
\firstname{Finn} \lastname{Stokes}\inst{1} \and
\firstname{Anthony} \lastname{Thomas}\inst{1,3} \and
\firstname{Samuel} \lastname{Thomas}\inst{1} \and
\firstname{Jia-jun} \lastname{Wu}\inst{1} 
\fnsep\thanks{This research was undertaken with the assistance of resources at the NCI National
  Facility in Canberra, the iVEC facilities at the Pawsey Centre and the Phoenix GPU cluster at the
  University of Adelaide, Australia.  These resources were provided through the National
  Computational Merit Allocation Scheme, supported by the Australian Government, and the University
  of Adelaide through their support of the NCI Partner Share and the Phoenix GPU cluster.
  This research was supported by the Australian Research Council through the ARC Centre of
  Excellence for Particle Physics at the Terascale (CE110001104) and through Grants
  No.\ DP151103101 (A.W.T.), DP150103164, LE120100181 and LE160100051 (D.B.L.).}
}
\institute{%
Special Research Centre for the Subatomic Structure of Matter (CSSM),
Department of Physics, University of Adelaide, Adelaide, South Australia 5005, Australia
\and
School of Physical Science and Technology, Lanzhou University, Lanzhou 730000, China
\and
ARC Centre of Excellence for Particle Physics at the Terascale
(CoEPP),
\\Department of Physics, University of Adelaide, Adelaide, South Australia 5005, Australia
}
\abstract{%
The structure of the ground state nucleon and its finite-volume excitations are examined from three
different perspectives.  
Using new techniques to extract the relativistic components of the nucleon wave function, the node
structure of both the upper and lower components of the nucleon wave function are illustrated.  A
non-trivial role for gluonic components is manifest.  
In the second approach, the parity-expanded variational analysis (PEVA) technique is utilised to
isolate states at finite momenta, enabling a novel examination of the electric and magnetic form
factors of nucleon excitations.  Here the magnetic form factors of low-lying odd-parity nucleons
are particularly interesting.  
Finally, the structure of the nucleon spectrum is examined in a Hamiltonian effective field theory
analysis incorporating recent lattice-QCD determinations of low-lying two-particle scattering-state
energies in the finite volume.  The Roper resonance of Nature is observed to originate from
multi-particle coupled-channel interactions while the first radial excitation of the nucleon sits
much higher at approximately 1.9 GeV.
}
\hfill ADP-17-38/T1044
\maketitle

%
\section{Introduction}
\label{sec:intro}

In this brief report, the structure of the nucleon and its finite-volume excitations are examined
from three different perspectives.  
The results are based on the PACS-CS $2 + 1$ flavour dynamical-fermion
configurations~\cite{Aoki:2008sm} made available through the ILDG~\cite{Beckett:2009cb}.
 
In Sec.~\ref{sec:WF} results from a new Lattice QCD examination
of the relativistic components of the nucleon's wave function are reported.  Both the ground state
and the first radial excitation of the nucleon are illustrated.
In Sec.~\ref{sec:EMFF}, first calculations of the electromagnetic form factors of the low-lying
odd-parity nucleon excitations are presented.  Here the Parity-Expanded Variational Analysis (PEVA)
technique is central to the isolation of states at finite momenta.  By comparing with results from
conventional variational analyses, the essential nature of the PEVA technique is manifest.
Finally, Sec.~\ref{sec:RR} explores the relationship between the radial excitation of the nucleon
and the Roper resonance of Nature in Hamiltonian Effective Field Theory (HEFT).  Including a bare
quark-model state at approximately 2 GeV as a basis state in the effective field theory, we find a
small but nontrivial role for the quark model in the Roper-resonance regime.  However, the
predominant structure of the Roper resonance has its origin in the strong rescattering of $\pi N$,
$\pi \Delta$ and $\sigma N$ channels.

%
\section{Relativistic components of the nucleon's wave function}
\label{sec:WF}

To explore the extended nature of the nucleon's wave function, we take the standard approach of
gauge fixing the gluon fields to Landau gauge \cite{Bonnet:1999mj,Bonnet:2002ih}.  This enables the
use of non-local interpolating fields where the quark field operators act at spatially separated
points within a time slice of the lattice.
To this end the baryon annihilation operator is generalised to the $u$-quark symmetric form
\begin{equation}
\epsilon^{abc}
\left ( {u}^{Ta}(\vec x + {\vec s}, t)\, C{\gamma_5}\,{d}^b(\vec x + {\vec y},
t) \right )\, {u}^{c}(\vec x - {\vec s}, t) + 
\epsilon^{abc}
\left ( {u}^{Ta}(\vec x - {\vec s}, t)\, C{\gamma_5}\,{d}^b(\vec x + {\vec y},
t) \right )\, {u}^{c}(\vec x + {\vec s}, t)  \, .
\end{equation}
%
The momentum of the state is projected via Fourier transform over $\vec x$ and the wave function is
obtained as a measure of the {\em overlap} of this operator with the state under consideration as a
function of the $d$-quark position ${\vec y}$ for fixed $u$-quark separation
$2{\vec s}$.
In this case, one obtains the wave function of the ${d}$ quark within the proton.
In this brief report, we consider the two ${u}$ quarks at rest at the origin,
${\vec s} = 0$.

To access isolated finite-volume energy eigenstates, $\alpha$, a correlation-matrix optimised
interpolating field ${\bar\phi}^{\alpha} = \sum_{i} u_{i}^{\alpha}\, {\bar\chi}_{i}$ is used at the
source.  Here $u_{i}^{\alpha}$ is the right eigenvector of the standard generalised eigenvalue
problem for local interpolating fields.  Four different levels of gauge-invariant Gaussian smearing
are considered as described in Ref.~\cite{Mahbub:2013ala} for $\chi_1$ and $\chi_2$.

The orbital angular momentum of the spatial wave function is constrained through the selection of
Dirac indices in the proton two-point function.  Denoting this $4\times 4$ Dirac matrix by
$G_{ba}(t)$, the source Dirac index $a$ selects the spin and parity of the state, $\alpha$, under
investigation.  Focusing on the zero-momentum positive-parity ground and radially-excited states,
the spin-up state, $\ket{\uparrow}$, has $a = 1$ and the spin-down state, $\ket{\downarrow}$,
corresponds to $a = 2$.

Dirac indices at the sink select one of the four components of the Dirac spinor wave function under
examination.  At zero momentum, $b = 1$ is $\ket{\uparrow}_{\rm upper}$, $b = 2$ is
$\ket{\downarrow}_{\rm upper}$, $b = 3$ is $\ket{\uparrow}_{\rm lower}$ and $b = 4$ is
$\ket{\downarrow}_{\rm lower}$.  The spatial wave function will compensate for spin and parity
mismatches between the source and sink.

Because the lower Dirac components have opposite parity, orbital angular momentum $\ell = 1$ is
required for the even parity ground and radially excited states.  Denoting the spatial wave function
by $\braket{\vec r \,|\, \alpha, \ell, m_\ell \,}$, and noting the gauge-invariant Gaussian smeared
source, ${\bar\phi}^{\alpha}$, has $\braket{\vec r \,|\, \alpha, 0, 0 \,}$, the matrix element
$G_{31}$ describes the source $\to$ sink transition
\begin{equation}
\braket{\vec r \,|\, \alpha, 0, 0 \,}\: \ket{\uparrow}_{\rm upper}
  \to 
 \ket{\uparrow}_{\rm lower}\, \braket{\vec r \,|\, \alpha, 1, 0 \,} \, .
\end{equation}
Similarly, matrix element $G_{42}$ describes the transition
\begin{equation}
\braket{\vec r \,|\, \alpha, 0, 0 \,}\: \ket{\downarrow}_{\rm upper}
  \to 
  \ket{\downarrow}_{\rm lower}\, \braket{\vec r \,|\, \alpha, 1, 0 \,} \, .
\end{equation}
In both cases, the $\ell=1$, $m_\ell = 0$ spatial wave function is resolved.

\begin{figure}[tb] 
  \centering
  \includegraphics[width=0.48\textwidth]{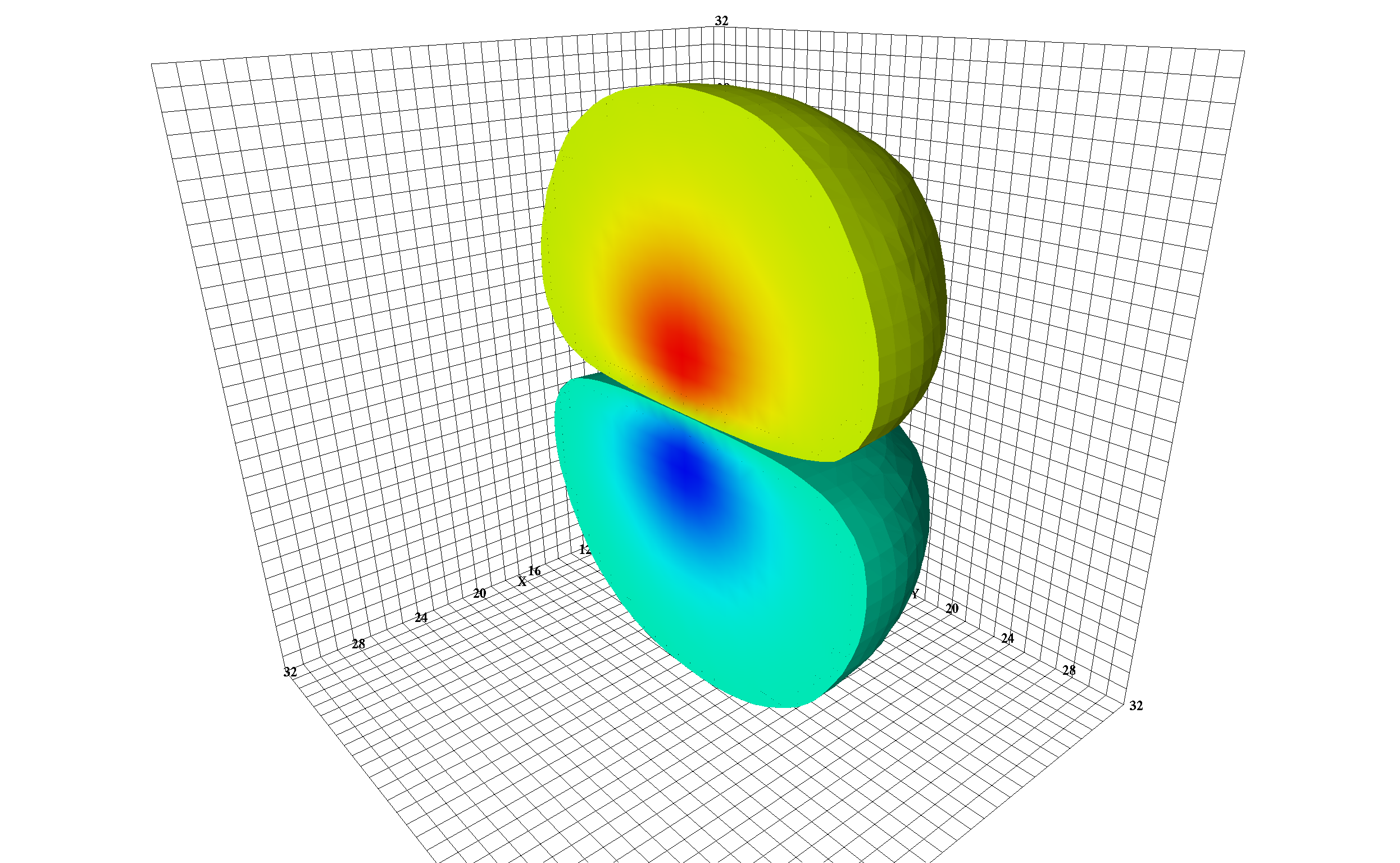}
  \includegraphics[width=0.48\textwidth]{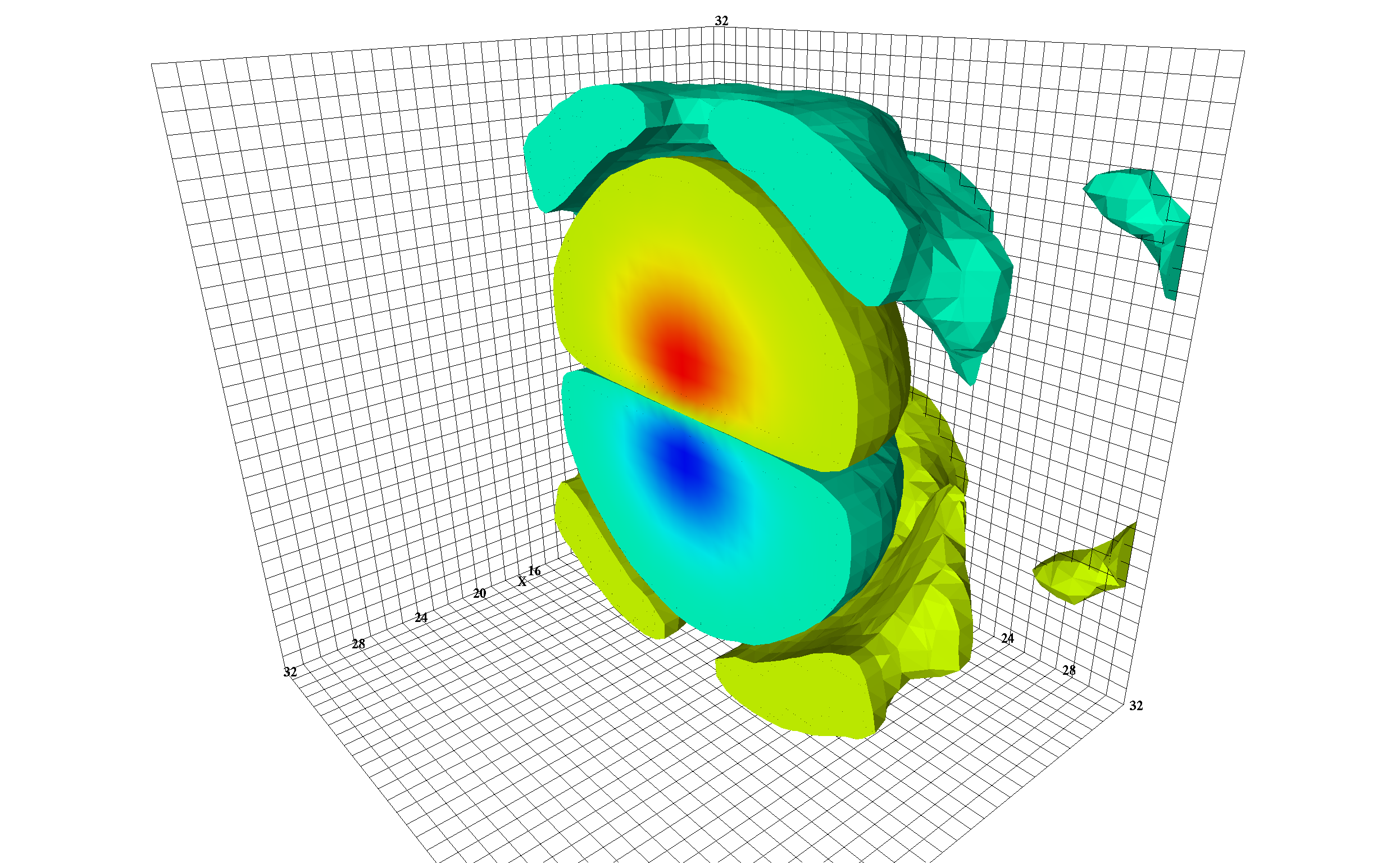}
  \caption{Ground (left) and first excited (right) lower-relativistic-component wave
    functions $\braket{\vec r \,|\, \ell, m_\ell \,} = \braket{\vec r \,|\, 1,0\,}$ for the $d$
    quark in the proton at light quark masses providing $m_\pi = 156$ MeV.  Amplitudes are
    illustrated with the sign of the wave function indicated by red-yellow volume rendering for
    positive values and blue-cyan volume rendering for negative values.  }
  \vspace{-12pt}
  \label{fig-1}
\end{figure}

Cubic-symmetry artefacts associated with the finite-periodic volume are manifest in the tails of
the wave functions.  Away from the boundaries, the anticipated spherical harmonics are observed.
Figure \ref{fig-1} illustrates this $d$-quark wave function for the proton ground state and first
radial excitation.  
Similar quality results are observed for the $\braket{\vec r \,|\, 1, \pm 1\,}$
spatial wave function associated with a spin flip.

%
The $G_{21}$ and $G_{12}$ matrix elements are particularly interesting.  The spin flip between the
source and sink requires a nontrivial value for $m_\ell$.  In addition, the upper-component $\to$
upper-component transition requires even parity in the spatial wave function.  In this case, $\ell
= 2$ is the first candidate.  However, $\ell + s = 2 + 1/2 \ne 1/2$ and therefore this matrix
element is expected to vanish on large volumes if quarks are the only source of angular momentum in
the nucleon.

\begin{figure}[tb]
  \centering
  \sidecaption
  \includegraphics[width=0.38\textwidth]{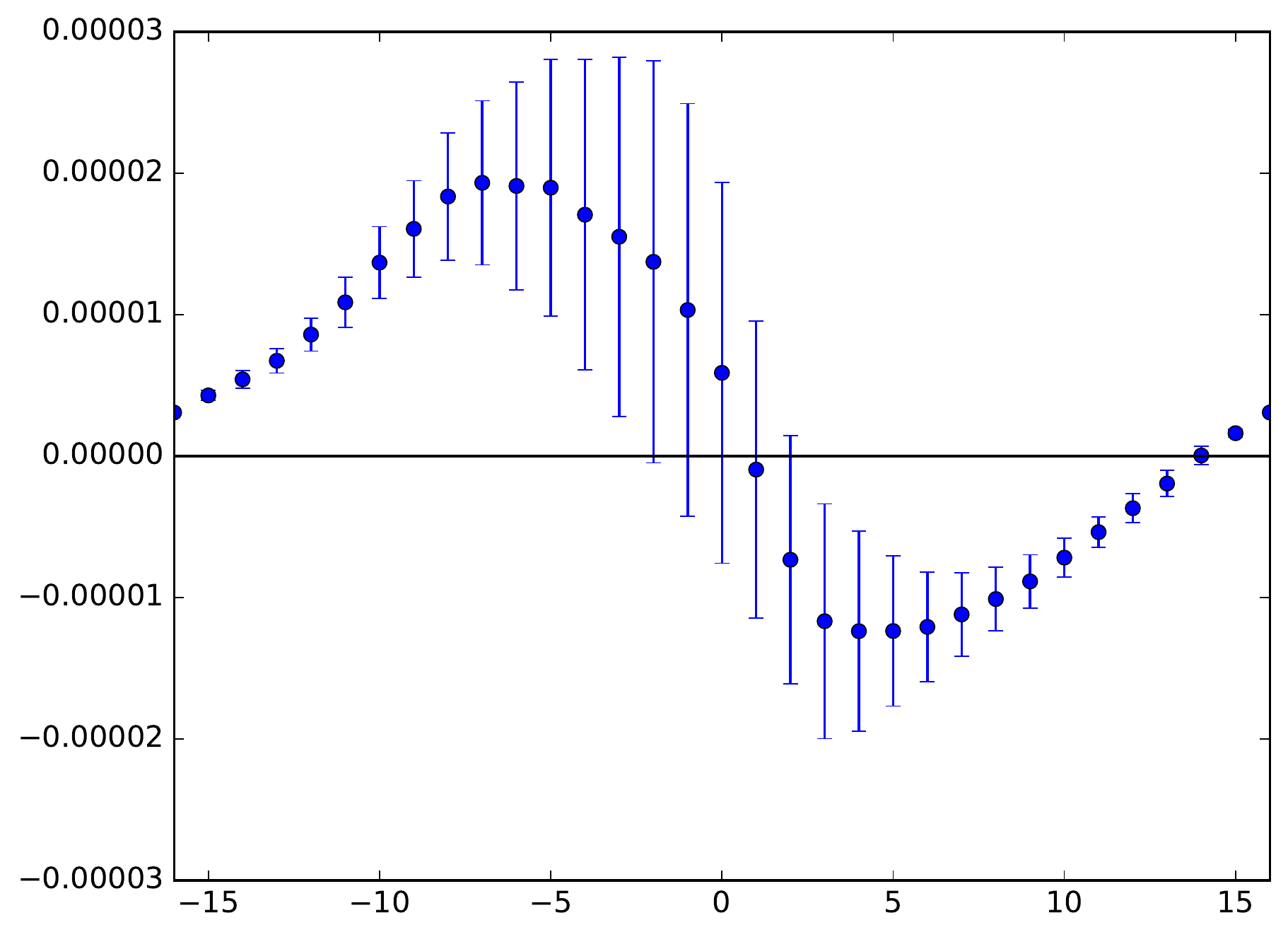}
  \caption{The imaginary part of the spatial wave function $\braket{\vec r \,|\, 1, 1 \,}_{\rm
      quark}$ obtained from the calculation of the $G_{21}$ matrix element.  The independent
    coordinate is along the $y$-axis of the lattice volume.  The y-axis illustrates the wave
    function amplitude relative to a probability normalization of 1.  The $p$-wave character in the
    quark sector demands a non-trivial $p$-wave contribution from the gluon sector to maintain the
    positive parity of the upper components of the wave function.  }
  \label{fig:gluons}
\end{figure}

However, lattice QCD calculations of this matrix element are nontrivial and therefore expose a
non-trivial role for gluonic degrees of freedom \cite{Bakry:2015csa} in carrying angular momentum
in the nucleon, a result well established through the measurement of generalized parton
distributions through deeply virtual Compton scattering experiments.
In the event $\ell = 1$ is the dominant contribution of the gluon field, the matrix element
$G_{21}$ is composed of
\begin{equation}
\braket{\vec r \,|\, 0, 0 \,}\: \ket{\uparrow}_{\rm upper}
  \to 
 \ket{\downarrow}_{\rm upper}\, \left (
\braket{\vec r \,|\, 1, 1 \,}_{\rm quark} \, \braket{\vec r \,|\, 1, 0 \,}_{\rm gluon}
+
\braket{\vec r \,|\, 1, 0 \,}_{\rm quark} \, \braket{\vec r \,|\, 1, 1 \,}_{\rm gluon}
\right ) \, .
\end{equation}
Symmetrisation can be used to expose the individual terms.  For example, symmetrisation along the
$z$ axis eliminates the antisymmetric $\braket{\vec r \,|\, 1, 0 \,}_{\rm quark}$ term, thus
isolating the $\braket{\vec r \,|\, 1, 1 \,}_{\rm quark}$ contribution.  While higher values of
$\ell$ are also admitted, Fig.~\ref{fig:gluons} illustrates that the $p$-wave orbital dominates.
We anticipate that appropriate symmetrisation/anti-symmetrisation of all the matrix elements
considered is likely to reveal nontrivial gluonic contributions.

%
\section{Electromagnetic form factors of nucleon excitations}
\label{sec:EMFF}

The isolation of excited baryons moving with finite momentum encounters difficulties with
cross-parity contaminations.  This problem is solved with the parity-expanded variational analysis
(PEVA) \cite{Menadue:2013kfi}.  We are now able to exploit this technique to conduct the
first examination of the electromagnetic form factors of nucleon excited states.
Our focus here is on the magnetic moments of low-lying odd-parity excitations of the proton, $p$, and
neutron, $n$.  Our interest is to determine if quark-model descriptions of their magnetic moments
are in agreement with the predictions of QCD.

One proceeds by expanding the basis of interpolating fields, $\chi^i$, to include both positive and
negative parity transformation properties.  One can then simultaneously isolate the finite-momentum
energy eigenstates of both parities.  Noting that terms in the unprojected correlation matrix have
the Dirac structure
\begin{equation}
\left (
\begin{array}{cc}
{E_{B^\pm}\!(\vect{p}) \pm m_{B^\pm}}  &{-\sigma_k p_k} \\
{\sigma_k p_k}                         &{-\left(E_{B^\pm}\!(\vect{p}) \mp m_{B^\pm} \right )}
\end{array}
\right ) \, ,
\end{equation}
we define PEVA projector
\begin{equation}
\proj = \frac{1}{4} (\identity + \gamma_4) (\identity - i \gamma_5 \gamma_k \hat{p}_k) \, ,
\end{equation}
and double the correlation matrix size through the consideration of $\chi_{\vect{p}}^{i} \definedby
\proj{}\, \chi^i$ which couples to positive parity states at zero momentum and
$\chi_{\vect{p}}^{i'} \definedby \proj{}\, \gamma_5\, \chi^i$ which couples to negative parity
states at zero momentum.  The generalised-eigenvalue equation is solved for the two-point functions
at finite momentum for Euclidean-time values $t_0 = 1$ after the source and $\Delta t = 3$ and the
corresponding eigenvectors are used to project the two and three-point functions to individual
eigenstates \cite{Liu:2015ktc,Mahbub:2013ala}.

Magnetic moments are estimated from the electromagnetic form factors under the assumption of a
similar $Q^2$ dependence for the electric and magnetic form factors from each quark sector
composing the baryon. Considering $G(Q^2)$ for single quark flavours of unit charge, we have
\begin{equation}
\mu_p = \frac{2}{3}\, 2\,  \frac{G_M^D(Q^2)}{G_E^D(Q^2)} - \frac{1}{3} \frac{G_M^S(Q^2)}{G_E^S(Q^2)}
\, , \quad \mbox{and} \quad
\mu_n = -\frac{1}{3}\, 2\,  \frac{G_M^D(Q^2)}{G_E^D(Q^2)} + \frac{2}{3} \frac{G_M^S(Q^2)}{G_E^S(Q^2)}
\, ,
\end{equation}
where $G^D$ denotes the doubly represented quark sector (eg.\ $u$ in $p$) and $G^S$ represents the singly
represented quark sector (eg.\ $d$ in $p$).  We find these ratios to be consistent for the various small
values of $Q^2$ considered through momentum boosts.

\begin{figure}[tb]
  \centering
  \includegraphics[width=0.8\textwidth]{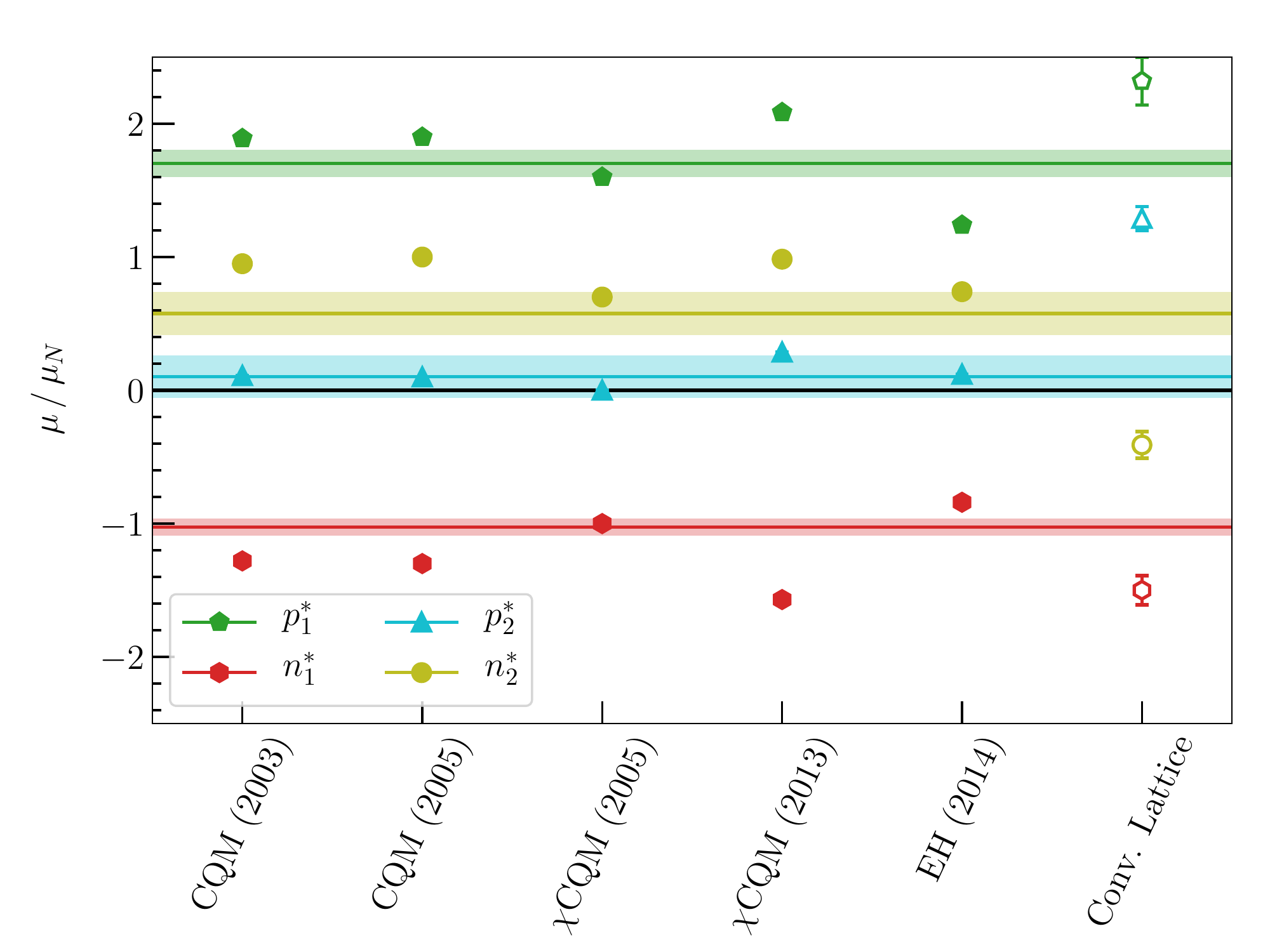}
  \vspace{-12pt}
  \caption{PEVA-based lattice QCD calculations of excited state
    magnetic moments (horizontal bands) are compared with quark-model
    expectations
    \cite{Chiang:2002ah,Liu:2005wg,Sharma:2013rka,Narodetskii:2013cxa}
    and lattice QCD results obtained from a conventional correlation
    matrix analysis where finite-momentum opposite-parity
    contaminations are not removed.  The finite-volume proton and
    neutron excitations associated with the $N^*$(1535) resonance are
    denoted $p_1^*$ and $n_1^*$, and excitations associated with the
    $N^*$(1650) resonance are denoted $p_2^*$ and $n_2^*$
    respectively.  }
  \vspace{-12pt}
  \label{fig:magMomComp}
\end{figure}

Figure \ref{fig:magMomComp} presents results for the finite-volume proton and neutron excitations
associated with the $N^*$(1535) resonance, denoted $p_1^*$ and $n_1^*$ respectively, and
excitations associated with the $N^*$(1650) resonance, denoted $p_2^*$ and $n_2^*$ respectively.
Our PEVA-based lattice QCD calculations of these four magnetic moments are illustrated as
horizontal bands reflecting the one-sigma uncertainties.

These lattice calculations are compared with a variety of quark models all of which display the
same qualitative pattern for the magnetic moments.  Quark models include the constituent quark
models of
Ref.~\cite{Chiang:2002ah} as CQM(2003),
Ref.~\cite{Liu:2005wg} as CQM(2005) and its extension to include meson-cloud
degrees of freedom as $\chi$CQM(2005), the chiral constituent quark
model of
Ref.~\cite{Sharma:2013rka} as $\chi$CQM(2013), and the effective Hamiltonian
approach of
Ref.~\cite{Narodetskii:2013cxa} as EH(2014).
The agreement of the chiral constituent quark model, $\chi$CQM(2005), is remarkable, indicating
these lattice-QCD states are described well by a quark model incorporating meson-cloud degrees of
freedom which act to suppress the magnitudes of the constituent quark moments.  

The lattice QCD results obtained from a conventional correlation matrix analysis, where
finite-momentum opposite-parity contaminations are not removed, differ significantly from the
PEVA-based analysis.  This observation demonstrates the necessity of using the PEVA technique in
all lattice QCD investigations of excited states.

These results provide a significant step forwards in understanding the complex emergent phenomena
of QCD.  It will be fascinating to discover if a constituent quark description can continue to
describe the nature of these states as the quark masses are reduced towards the values realised in
Nature \cite{Leinweber:2002qb,Hall:2013oga,Wang:2013cfp}.  In the case of the $\Lambda(1405)$ the
low-lying \cite{Menadue:2011pd,Leinweber:1989hh} state was observed to rearrange itself into a
molecular $\overline K N$ meson-baryon bound state \cite{Hall:2014uca,Hall:2014gqa}.  There the
vanishing of the strange-quark magnetic form factor in lattice QCD is consistent with the
strange-quark's confinement within a spin-zero $K$ meson having zero angular momentum about the
nucleon \cite{Hall:2014uca}.  Similarly, the light-quark contributions are observed to be governed
by the nucleon magnetic form factors \cite{Hall:2016kou,Hall:2015cua}.  At sufficiently light quark
masses, a molecular meson-baryon composition for the $N^*$ states observed on the lattice may
become an important component.

%
\section{The Roper resonance is a multi-particle coupled-channel state}
\label{sec:RR}

Every lattice QCD collaboration using local three-quark interpolating fields in a correlation
matrix analysis has failed to observe a low-lying positive-parity nucleon excitation in the
vicinity of the Roper resonance \cite{Mahbub:2010rm, Edwards:2011jj, Mahbub:2012ri, Engel:2013ig,
  Mahbub:2013ala, Roberts:2013ipa, Roberts:2013oea, Alexandrou:2014mka,
  Kiratidis:2015vpa,Liu:2016uzk}.  Motivated by the molecular meson-baryon bound state discovered
in the $\Lambda(1405)$ channel \cite{Hall:2014uca,Hall:2014gqa}, an extensive search for a
molecular meson-baryon description of the Roper was carried out in Ref.~\cite{Kiratidis:2016hda}.
There, local $\pi N$, $\sigma N$ and $a_0 N$ interpolators were explored in a variety of
correlation matrix analyses in an attempt to reveal a nucleon excitation in the Roper regime.
However, no new low-lying states were observed.

These results were confirmed in an independent manner by Lang {\it et al.}~\cite{Lang:2016hnn}.  In
this work, five-quark {\it non-local} interpolating fields with the momenta of each hadron
projected in the interpolator were considered.  Their calculations incorporate the $\pi N$ and
$\sigma N$ channels and two new low-lying scattering states were observed.  The composition of these
states is consistent with the scattering-state interpolators used to excite them from the vacuum.
Their third excited state is three-quark dominated and agrees with the results of
Ref.~\cite{Liu:2016uzk}.

With the first quark-model radial excitation of the nucleon sitting at 1.9(1) GeV in lattice QCD
\cite{Liu:2016uzk, Roberts:2013ipa, Roberts:2013oea}, one might wonder if it possible to reconcile
the large masses of localised quark-model like states with the observed low-lying position of the
Roper resonance.  This question was addressed in the Hamiltonian Effective Field Theory (HEFT)
analysis of Ref.~\cite{Wu:2017qve}.  HEFT \cite{Hall:2013qba} is an extension of chiral
perturbation theory that incorporates the L\"uscher relation connecting the energy levels observed
in finite volume to the scattering phase shifts \cite{Wu:2014vma}.  In the power-counting regime
HEFT reproduces the expansion of chiral perturbation theory for ground state phenomena
\cite{Young:2002ib}.

\begin{figure}[t]
\begin{center}
\includegraphics[width=0.72\textwidth]{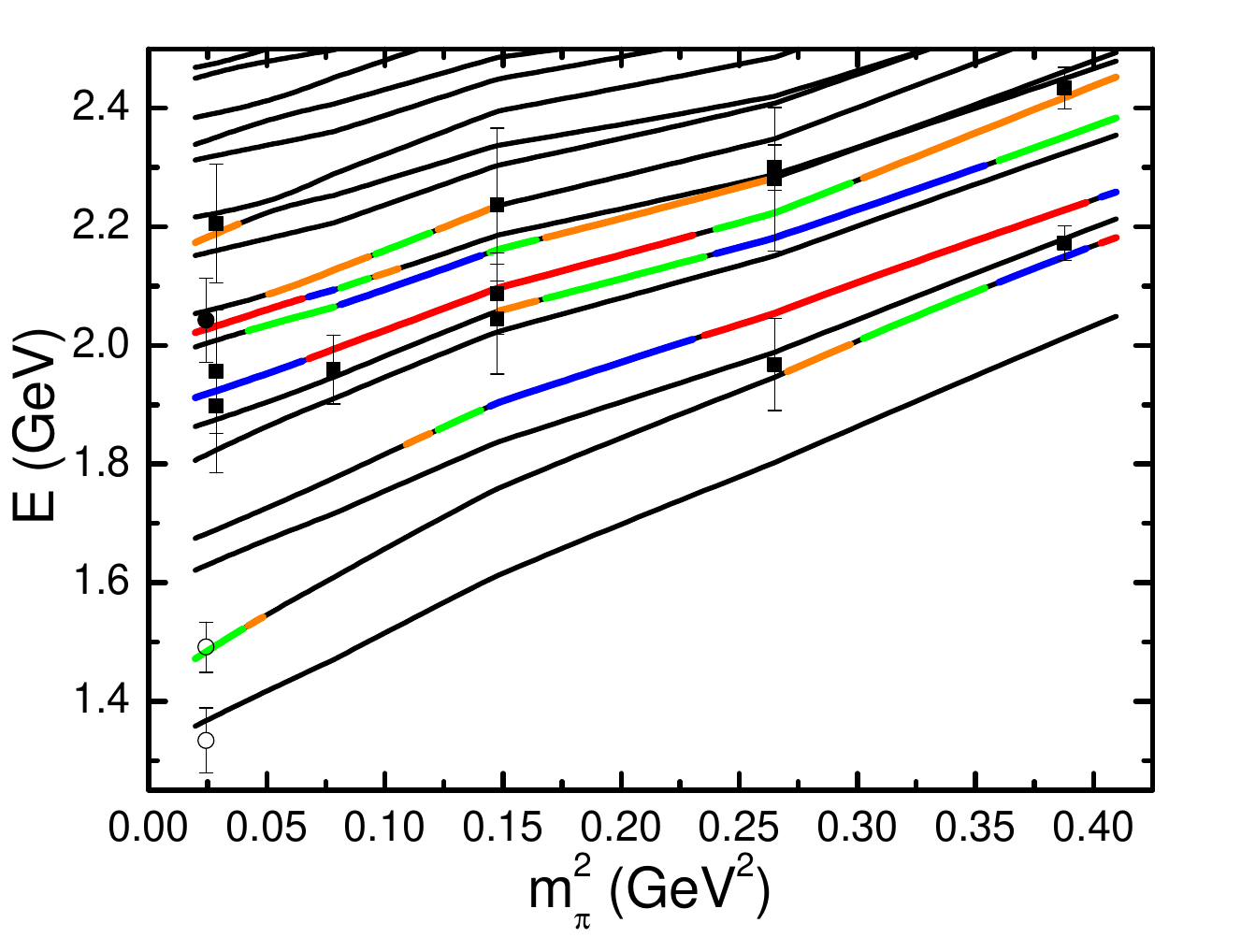}
\vspace{-12pt}
\caption{ The finite-volume HEFT spectrum of Ref.~\cite{Wu:2017qve} is compared with contemporary
  lattice QCD results.  Here the bare quark-model-like basis state has a mass of 2.0 GeV.  The CSSM
  lattice-QCD results \cite{Liu:2016uzk} are indicated by square symbols and circles denote the
  more recent lattice results from Lang {\it et al.}\ \cite{Lang:2016hnn}.  Solid symbols indicate
  states dominated by local three-quark operators and open symbols indicate states dominated by
  non-local momentum-projected five-quark operators.  The colours red, blue, green and orange are
  used to indicate the relative contributions of the bare quark-model-like basis state in the
  eigenstate, with red being the largest contribution.  }
  \vspace{-30pt}
  \label{fg:nuclSpec}
\end{center}
\end{figure}

Constraining the coupled $\pi N$, $\pi \Delta$ and $\sigma N$ channels to experimental scattering
data and ensuring the finite-volume states dominated by the bare quark-model-like basis state are
associated with localised states observed in lattice QCD produces the finite-volume spectrum of
Fig.~\ref{fg:nuclSpec}.  While the mass of the bare quark-model-like basis state is $2.0$ GeV, this
fit provides a pole in the Roper resonance regime at $1393 - i\, 167$ MeV.

Within the context of a constituent quark model, a mass of 2 GeV for the first radial excitation is
natural.  With the first negative-parity excitation of the nucleon, the $N^\star(1535)$, almost 600
MeV above the nucleon, expectations -- based upon the harmonic oscillator model which has enjoyed
success in hadron spectroscopy -- suggest that the first positive-parity excited state
should occur around 2 GeV.  

The description of both the experimental scattering data \cite{Wu:2017qve} and the lattice data
presented in Fig.~\ref{fg:nuclSpec} is excellent.  All localised lattice QCD states are associated
with a HEFT spectral line whose composition includes a large quark-model-like basis state
component.  Similarly, HEFT accurately predicts the positions of the scattering states observed in
lattice QCD by Lang {\it et al.} \cite{Lang:2016hnn} as well as their compositions.  It's
particularly interesting to note that the second excitation observed by Lang {\it et al.} at
$\simeq 1.5$ GeV -- created with the momentum-projected $\sigma N$ interpolating field -- has a
small 2 GeV quark-model-like basis-state component in HEFT, which may become the dominant component
in large $Q^2$ transition form factors \cite{Wilson:2011aa}.  However, the predominant structure of
the Roper resonance has its origin in the strong rescattering of $\pi N$, $\pi \Delta$ and $\sigma
N$ channels.
\vspace{-6pt}

%
\section{Conclusions}
\label{sec:concl}

The relativistic components of the ground-state nucleon and its first localised excitation have been
explored in lattice QCD.  The node structure in both the large and small components of the
relativistic wave function is consistent with the state observed at 1.9(1) GeV being the first
radial excitation of the nucleon.

The Parity Expanded Variational Analysis (PEVA) Technique enables an exploration of the
electromagnetic form factors of nucleon excitations.  The PEVA technique is
vital to the accurate determination of excited state form factors as the conventional variational
analysis admits large opposite-parity contaminations.  The merit of the PEVA technique in 
precision matrix-element calculations of the ground state nucleon is also apparent in the lattice
QCD three-point correlation functions.

We have discovered the two localised low-lying odd-parity nucleon states of the lattice QCD nucleon
spectrum are unlikely to be finite-volume manifestations of a single resonance.  They have magnetic
moments consistent with quark model expectations
for the $N^*(1535)$ and $N^*(1650)$.  Future studies will map the quark mass dependence of these
electromagnetic form factors and further unveil the structure of these nucleon excitations.

Finally, the relationship between the radial excitation of the nucleon and the Roper resonance of
Nature is now understood.  The quark-model-like basis state at approximately 2 GeV makes a small
but nontrivial contribution to the finite volume state observed in the regime of the Roper
resonance.  However, the predominant structure of the Roper resonance has its origin in the strong
rescattering of $\pi N$, $\pi \Delta$ and $\sigma N$ channels.
\vspace{-6pt}

\bibliography{StructNuclEx}

\end{document}